\documentclass[submission,11pt,authoryear, nonatbib]{elsarticle}

\usepackage{booktabs}
\usepackage[export]{adjustbox}
\usepackage{graphicx}
\usepackage{graphbox}
\usepackage{colortbl}
\usepackage{subcaption}
\usepackage{enumitem}
\usepackage[natbibapa]{apacite}

\usepackage{amsmath}
\usepackage{amssymb}
\usepackage{lineno}
\usepackage[hidelinks]{hyperref}
\usepackage[utf8]{inputenc}
\usepackage{newunicodechar}
\usepackage[utf8]{inputenc}
\usepackage{textcomp}

\journal{Body Image}

\begin{document}

\begin{frontmatter}

\title{Leveraging Machine Learning to Identify Gendered Stereotypes and Body Image Concerns on Diet and Fitness Online Forums}



\author[1]{Minh Duc Chu\corref{cor1}\fnref{equal}}
\ead{mhchu@usc.edu}
\author[2]{Cinthia S\'anchez\fnref{equal}}
\ead{mabel.sanchez.macias@gmail.com}
\author[1]{Zihao He}
\ead{zihaoh@usc.edu}
\author[1]{Rebecca Dorn}
\ead{rdorn@usc.edu}
\author[3]{Stuart B Murray\fnref{ci}}
\ead{SBMurray@mednet.ucla.edu}
\author[1]{Kristina Lerman\fnref{ci}}
\ead{lerman@isi.edu}

\address[1]{Information Sciences Institute, University of Southern California, Marina del Rey, CA 90292, USA}
\address[2]{Department of Computer Science, University of Chile, Santiago, Chile}
\address[3]{Department of Psychiatry and Biobehavioral Sciences, University of California, Los Angeles, CA 90095, USA}

\fntext[equal]{These authors contributed equally to this work.}



\begin{abstract}
\textcolor{red}{[\textbf{Warning: This paper discusses eating disorders and body dysmorphia, which some may find distressing.}]}\\

\noindent The pervasive expectations about ideal body types in Western society can lead to body image concerns, dissatisfaction, and in extreme cases, eating disorders and other psychopathologies related to body image. While previous research has focused on online pro-anorexia communities glorifying the ``thin ideal'', less attention has been given to the broader spectrum of body image concerns or how emerging disorders like muscle dysmorphia (``bigorexia'') present in online platforms. To address this gap, we analyze 46 Reddit forums related to diet, fitness, and mental health. We mapped these communities along gender and body ideal dimensions, revealing distinct patterns of emotional expression and community support. Feminine-oriented communities, especially those endorsing the thin ideal, expressed higher levels of negative emotions and received caring comments in response. In contrast, muscular ideal communities displayed less negativity, regardless of gender orientation, but received aggressive compliments in response, marked by admiration and toxicity. Mental health discussions aligned more with thin ideal, feminine-leaning spaces. By uncovering these gendered emotional dynamics, our findings can inform the development of moderation strategies that foster supportive interactions while reducing exposure to harmful content.
\end{abstract}


\end{frontmatter}


\section{Introduction}
Social norms shape what is considered attractive and healthy in a society, and also how individuals perceive and evaluate their own bodies. In Western cultures, this often differs for men and women. From an early age, women are socialized to pursue the ``thin ideal'', a body type characterized by slenderness and low body fat, while men are expected to be strong and visibly muscular, as embodied by the ``muscular ideal''~\citep{Calogero2010,murnen2012body}. These cultural expectations reflect broader gender norms that associate women’s value with attractiveness and men’s with strength and control. Internalizing these body ideals can generate body image concerns and body dissatisfaction, which, in extreme cases, contribute to eating disorders and related psychopathologies~\citep{mclean2019body}. 
However, in contrast to the extensive literature on thinness and the thin ideal, the ways in which individuals internalize the muscular ideal, and how they express distress or seek support around it, remain poorly understood.

\subsection{Sociocultural Mechanisms of Body Image Disturbance}

The mechanisms through which the gendered body ideals influence body dissatisfaction and disordered eating have been described in several sociocultural theories. The Tripartite Influence Model~\citep{Keery2004} identifies peers, parents, and media as key transmitters of appearance norms that promote body dissatisfaction through internalization of body ideals and appearance-based social comparison. Objectification Theory~\citep{fredrickson1997objectification} postulates that both male and female bodies are socially constructed as objects of visual and moral evaluation, fostering social pressures to be thin and body shame. In parallel, masculinity frameworks~\citep{griffiths2014young,murray2016go} describe how muscularity and self-discipline function as performances of masculine identity, promoting rigid food rules and exercise behaviors among men. These theories collectively establish body image concerns as gendered, not because of inherent differences between men and women, but because sociocultural systems prescribe \textit{fundamentally different} forms of embodiment and emotional expression across genders.


\subsection{Body Ideals and Social Media}

Body dissatisfaction long predates social media~\citep{brumberg2000fasting}. However, social media platforms, where users create, share, consume and evaluate appearance related content have created new risks by amplifying exposure to idealized bodies and social comparison processes, thereby amplifying mechanisms which promote body dissatisfaction and concomitant disordered eating~\citep{Saiphoo2019,choukas2022perfectstorm}. Online platforms link people to digitally modified images of objectified others, with visible metrics of perceived desirability (e.g., Likes, Comments), sometimes without users even searching for such content, which may amplify the impact of media and peer influences. Research has documented pro-anorexia and weight-loss communities that provide peer influences that normalize restrictive behaviors and glorify the thin ideal~\citep{juarascio2010pro,oksanen2016proanorexia,wang2018social,lerman2025safe}. Engagement in such spaces is associated with increased body dissatisfaction~\citep{de2016adolescents,pedalino2022instagram} and disordered eating~\citep{griffiths2024does}; however, it can also provide social support and belonging for those who feel stigmatized or misunderstood offline~\citep{YeshuaKatz2013stigma,chancellor2016recovery}. 


In contrast to the thinness-oriented communities, \textit{muscularity-oriented} communities have received relatively little scholarly attention. 
Individuals with muscularity-oriented disordered eating, sometimes referred to as muscle dysmorphia or ``bigorexia,'' frequently valorize compulsive exercise, restrictive diets, and the use of performance-enhancing drugs~\citep{murray2016go,Ganson2022cheatmeals,quiniones2019embracing}. This condition is associated with heightened psychiatric morbidity, substance misuse, and suicidality \citep{lavender2017men}. Despite these risks, little is known about how individuals experiencing muscularity concerns express distress, seek support, or engage with peers online.

\subsection{Gendered Emotion and Support in Online Communities}

Social and emotional expression in digital contexts is also shaped by gender norms. Women are more likely to express vulnerability and seek social connection, whereas men are often socialized to display emotional restraint and self-sufficiency~ \citep{brody1997gender,chaplin2015gender}. These patterns may manifest differently across thinness- and muscularity-oriented communities, influencing both the emotional tone of discussions and the nature of social support exchanged. Understanding these gendered emotional dynamics can illuminate how online environments reinforce or challenge sociocultural expectations about gender and body ideals.

\subsection{This Study}

This study examines how body ideals are expressed within online diet, fitness communities and the role gender plays in these expressions, as well as community engagement. We focus on  Reddit, a popular social media platform hosting a variety of discussion forums, which integrates peer interaction and media influence. We analyze 46 discussion forums on topics related to diet, fitness, body image, and mental health. We use state-of-the-art machine learning tools like graph embeddings to analyze community membership and transformer-based language models to classify emotions from text. Our work answers the following research questions: {(1)} How do online diet and fitness communities reflect the internalization of culturally gendered body ideals along the thin–muscular spectrum? How common are counter-stereotypical alignments, such as women driven by muscularity concerns?
(2) How do emotional expressions within these communities reveal gendered patterns of appearance-based social comparison? What are the differences in the tone and valence of emotions (e.g., sadness, admiration, disapproval) across body-ideal communities? 
(3) How do peer interactions differ in thinness- and muscularity-oriented online communities? How do gendered norms shape peer support,  visibility of psychological distress, and access to mental health support across body-ideal communities?

Our study demonstrates the potential of artificial intelligence and machine learning tools to systematically examine large-scale online discussions, revealing emergent psychological and social dynamics that can inform and extend existing theories of body image and body dissatisfaction. Computational analyses of online discourse can complement traditional methods by revealing how sociocultural pressures and gender norms manifest in contemporary digital contexts at scale.
Critically, integrating psychological theory via computational constructs can inform the development of more sensitive and robust moderation strategies for online communities. Models that account for emotional tone, gendered expression, and patterns of social support can help identify when discussions shift from supportive to potentially harmful, without pathologizing distress. Such interdisciplinary approaches move beyond detection of ``toxic'' content toward a more nuanced understanding of context, empathy, and community norms.

\section{Methods}
We apply machine learning techniques to large-scale textual data collected from Reddit. Our approach leverages computational methods to assess emotional tone, analyze content structure, and social structure and dynamics of discussions across a diverse set of online communities.  This combination of large-scale data collection and automated analysis allows us to systematically examine how users express emotion and psychological distress, seek and provide social support, and engage with thinness- and muscularity-oriented body ideals in online spaces.

\subsection{Data}
\label{sec:data}

We collected data from Reddit, a widely used social media platform with active communities (\textit{subreddits}) spanning diverse topics \citep{hofmann2022reddit}. To identify relevant communities, we first reviewed existing literature to generate keywords related to our focus areas, such as \textit{keto, gym, shredded, weightloss}, and used these keywords to locate seed subreddits on fitness, diet, and eating disorders. Leveraging Reddit’s ``Related Subs'' feature, we expanded our collection by identifying and manually verifying additional subreddits based on their descriptions and sample posts. The search process concluded once no further relevant or active subreddits were identified.

We applied the Louvain algorithm (details in $\S$\ref{sec:redditmention}) to the manually curated seed set of subreddits, identifying additional subreddits frequently mentioned by users in the seed communities. This approach enhanced coverage and reduced the likelihood of missing relevant communities. The process yielded 54 subreddits focused on fitness, diet, eating disorders, and related mental health topics. A manual verification was performed on the expanded set to ensure relevance. The dataset includes submissions created between January 2019 and November 2023, acquired via Academic Torrents\footnote{https://academictorrents.com/}, which aggregates Reddit data through the Pushshift API \citep{Baumgartner_Zannettou_Keegan_Squire_Blackburn_2020}.

\begin{table}[ht]
\small
\begin{tabular}{p{0.95\linewidth}}
\toprule
\multicolumn{1}{c}{\textbf{Subreddit (r/)}}             \\ \hline
                             
1200isplenty, amiugly, \textbf{AnorexiaNervosa}, ARFID, BDDvent, BingeEatingDisorder, \textbf{BodyDysmorphia}, \textbf{bodybuilding}, Bodyweightfitness, \textbf{Brogress}, \textbf{BulkOrCut}, CICO, \textbf{eating\_disorders}, \textbf{EDanonymemes}, \textbf{EDAnonymous}, EdAnonymousAdults, \textbf{EatingDisorders}, fasting, fit, \textbf{Fitness}, FlexinLesbians, fuckeatingdisorders, \textbf{gainit}, \textbf{GettingShredded}, goodrestrictionfood, Instagramreality, \textbf{intermittentfasting}, \textbf{keto}, ketogains, ketorecipes, \textbf{loseit}, MadeOfStyrofoam, nattyorjuice, omad, powerbuilding, \textbf{progresspics}, safe\_food, ShittyRestrictionFood, steroids, SuicideWatch, Volumeeating, weightroom, xxfitness, xxketo, drunkorexia, bulimia
\\ \bottomrule
\end{tabular}
\caption{Subreddits collected in our data, with prefix ``r/'' removed. Subreddits used to seed data collection are marked in bold.}
\label{tab:subreddits}
\end{table}


For each of the 54 subreddits, we remove submissions and comments from \textit{AutoModerator}, a bot that allows subreddit moderators to automate certain moderation tasks, including automatic posts and replies. Additionally, we discard content generated by specific subreddit bots such as \textit{steroidsBot, bodybuildingbot, Anabotlics, WeightroomBot} and \textit{EDAnonymous\_Bot} as these bots generated a large proportion of the content. Finally, we exclude deleted and duplicate content in both submissions and comments from our data. We then filter out subreddits with fewer than 500 submissions, leading to a final list of 46 subreddits, as shown in Table \ref{tab:subreddits}.

By nature, the number of submissions is outweighed by comments in any subreddit. To counteract the imbalance of post type, we employ random sampling. For each subreddit, we randomly sample at most 5,000 submissions and 5,000 comments. This leads to a total of 178,272 submissions, 218,139 comments, and 212,529 unique users.



\subsection{Subreddit Mention Network}
\label{sec:redditmention}

We construct a directed network of subreddit mentions. Nodes $V$ are subreddits, and a directed edge $u \rightarrow v$ is present when a post or comment in subreddit $u$ contains a clickable mention \texttt{r/v} in the title or body. For example, a user in \texttt{r/EatingDisorders} might write, ``If your main goal is weight loss, try \texttt{r/loseit},'' which creates a link from \texttt{r/EatingDisorders} to \texttt{r/loseit}. These links do not imply that the same person posts in both communities; they capture how communities direct users toward other communities for the purpose of getting information or support~\cite{krohn2022subreddit}. Mentions are extracted with the pattern \texttt{r/{subreddit\_name}}. Multiple mentions of the same target within one post are counted once, and the edge weight is the total number of posts and comments in which $u$ mentions $v$. We ignore posts generated by \textit{AutoModerator} and subreddit-specific bots, and we restrict the analysis to subreddits that are mentioned at least ten times so that each node has sufficient engagement. 

\subsection{Mapping Subreddits onto Social Dimensions} 
\label{sec:mapping_method}
Alternatively, we can examine patterns of user co-activity in different forums or communities to understand the organization of Reddit along social dimensions, such as body ideal or gender, using the community embedding method of \citet{waller2021quantifying}. This method 
uses neural embeddings to map Reddit communities into a high-dimensional space based on shared user activity and then projects each community's vector onto a social dimension defined by exemplar communities.
We begin with a user-community bipartite graph $G=(U,C,E)$, where $U$ is the users and $C$ is the subreddits. A link $e: u \leftrightarrow c$ exists between a user $u$ and a community $c$ if the user has commented or posted in the subreddit, with edge weight equal to the frequency of engagement. Communities that share many active users will sit close together in the resulting embedding space, so this graph captures overlap in participation rather than explicit mentions.

\begin{sloppypar}
The next step is to identify several seed pairs $P=\{(c^a_i, c^b_i)\}$, where each pair consists of communities that are similar in topic and norms but differ in their placement along a targeted social dimension. For example, \texttt{r/loseit} and \texttt{r/gainit} are both peer-support forums about weight change, but one is oriented toward weight loss and the other toward weight gain. Following \citet{waller2021quantifying}, we use these pairs as heuristic anchors rather than fixed ground truth: they are evidence-guided estimates of the poles that make the axis interpretable. For the thin--muscular body ideal dimension, we select four such pairs, [\texttt{r/loseit}, \texttt{r/gainit}], [\texttt{r/intermittentfasting}, \texttt{r/steroids}], [\texttt{r/AnorexiaNervosa}, \texttt{r/GettingShredded}], and [\texttt{r/Instagramreality}, \texttt{r/nattyorjuice}], by inspecting subreddit descriptions and highly upvoted posts and confirming that the language in each forum consistently aligns with the intended pole. For the gender dimension, we reuse the gender axis and subreddit scores estimated by \citet{waller2021quantifying} on a much larger Reddit graph. The resulting axes place communities along the spectrum of discourse style; they are not labels of individual users.
\end{sloppypar}

Following \citet{waller2021quantifying}, we then generate community embeddings by vectorizing subreddits in $G$ using \texttt{node2vec} \citep{grover2016node2vec}---a widely used algorithm that creates learned vector representations of nodes on a graph. 
We leverage these embeddings to create our axis $x$ representing the ideological/social spectrum by taking the average embeddings of communities on one side from the pair minus that on the other side, as $x = \frac{1}{n} \sum_{i=1}^n (e(c^a_i) - e(c^b_i))$, where $e(\cdot)$ represents the embedding of the community.  We project the community onto the axis $x$ by computing the cosine similarity between the community embedding and the axis. This process assigns each community a score on the dimension, indicating the similarity of its membership to the seeds at either end.

\subsection{Measuring Content Similarity}
Online communities' language captures their mindset and beliefs \citep{jiang2022communitylm, he2024reading, he2024community}. This enables us to use the text of the posts to measure the distance between communities in the semantic space. 
We encode the submissions of each subreddit using 
All MPNet Base V2\footnote{\url{https://huggingface.co/sentence-transformers/all-mpnet-base-v2}} sentence embedding model \citep{10.5555/3495724.3497138} finetuned on multiple datasets totaling more than 1 billion sentence pairs, including Reddit data, to capture sentence-level semantics. 

We approximate the textual and thematic 
similarity between subreddits using the 
Fréchet Inception Distance (FID) \citep{10.5555/3295222.3295408} that was originally developed to evaluate the quality of generated images but has been adapted to measure similarity in text corpora \citep{kour-etal-2022-measuring}. 
We identify groups of subreddits with similar content using agglomerative hierarchical clustering, applying the average linkage method with the Euclidean distance metric. This approach builds clusters by iteratively merging the most similar subreddits based on the FID metric, with average linkage calculating similarity as the average distance between all points in each cluster pair. 
We select the number of clusters that maximizes the silhouette score, which measures clustering quality by comparing how similar each point is to its cluster versus others, with higher scores indicating more distinct clusters.

\subsection{Measuring Emotions and Toxicity}
We measure toxic language and discrete emotions expressed in the text of posts and comments using transformer-based models, as described below. 

\subsubsection{Emotions}
Emotional expression in peer-to-peer interactions and dysfunctional emotion regulation have been linked to heightened body image concerns \citep{cash2012cognitive, hughes2011emotion}. 
To measure emotions in text, we use a RoBERTa-based model \texttt{roberta-base-go\_emotions}\footnote{https://huggingface.co/SamLowe/roberta-base-go\_emotions}, which has been fine-tuned on the GoEmotions dataset \citep{demszky2020goemotions} for multilabel classification. GoEmotions, like our data, consists of Reddit comments labeled with 28 categories, including 27 emotions and a neutral category (no emotions expressed). This model produces a score between 0 and 1, representing the confidence for each emotion category, including neutrality. Similar transformer-based models trained on GoEmotions are popular for emotion detection in social media analysis \citep{alhuzali2021spanemo, shmueli2021detecting,burghardt2023socio}.
To avoid noise, we consider the top 75\% of submissions and comments with the highest emotional content, filtering out submissions and comments with a neutral score above the 75th percentile separately.

Because the classifier operates on text, its scores capture emotion as expressed in language rather than physiological arousal. Sentence-level transformer models use the full sentence and nearby context, which allows them to pick up many connotative and culture-specific uses of language, but they can still misread some expressions. Throughout, we therefore interpret emotion scores as describing discourse norms within communities, not as direct measures of how emotionally reactive individual men or women are.

\subsubsection{Toxicity}
To detect toxicity, we utilize the \texttt{Detoxify} library \citep{Detoxify}, which provides a real toxicity score ranging from 0 to 1, with 1 indicating high toxicity. Additionally, \texttt{Detoxify} returns scores for various specific types of toxicity, such as \textit{obscene}, \textit{threat}, \textit{insult}, and \textit{identity attack}. \texttt{Detoxify} is a robust, widely-recognized model for accurately detecting nuanced toxic language, making it ideal for analyzing online social interactions {\citep{biasesYouTube2024}}.
To avoid noise, we only consider the top 75\% of submissions and comments with the highest levels of toxicity, filtering out content with negligible toxicity levels that might have been misclassified (submissions and comments with a toxicity score below the 25th percentile are excluded separately).


\section{Results}
We apply the quantitative methods described above to the textual data from Reddit diet and fitness forums to study our research questions.

\subsection{Mapping Subreddits onto the Gender and Body Ideal Spectrum}
\label{sec:res_body_ideal_dim}
Subreddits are mapped to the thin-muscular ideal and the gender axes using the methodology outlined in Section \ref{sec:mapping_method}. 

\subsubsection{Body Ideal Axis}

\begin{figure}[ht]
  \centering
      \includegraphics[width=\columnwidth]{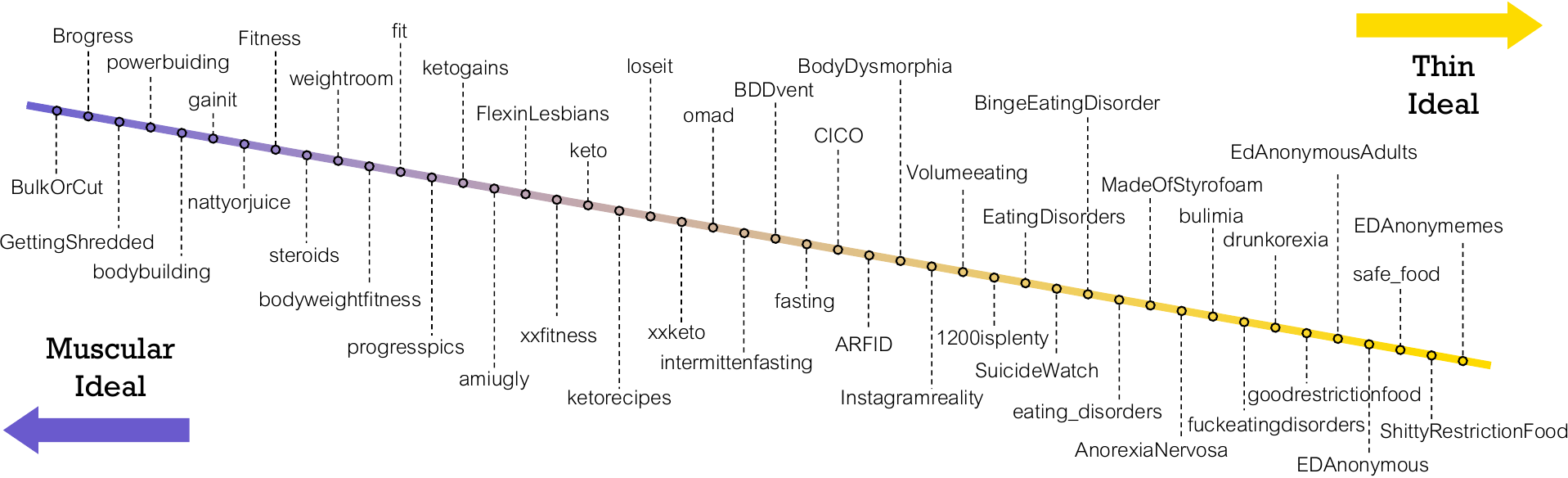}
  \caption{Ranking of subreddits along the muscular-thin ideal dimension, measured by cosine similarity. Subreddits on {the left} discuss the muscular ideal (e.g. \texttt{r/Getting\-Shredded}), while communities on the right promote the thin ideal (e.g. \texttt{r/AnorexiaNervosa}). Notably, mental health (non-eating-disorder) subreddits (e.g., \texttt{r/SuicideWatch}) are positioned near the Eating Disorder (ED) communities on the right side.}
  \label{fig:thin-musc-1d}
\end{figure}


Projecting the subreddits onto this axis yields scores representing each community's position along the body ideal spectrum. Higher scores indicate a stronger association with the thin ideal, while lower scores align more with the muscular ideal. Importantly, a community’s position reflects its association with this body ideal, not the identity of individual members. We manually validate the spectrum's accuracy by checking each forum’s description and reviewing a sample of user posts to ensure semantic alignment with the targeted body ideal.

Fig. \ref{fig:thin-musc-1d} positions the subreddits along the body ideal axis: those on the left emphasize the muscular ideal (e.g., \texttt{r/Getting-Shredded}), while those on the right promote the thin ideal (e.g., \texttt{r/AnorexiaNervosa}). 
Female-oriented fitness communities (e.g., \texttt{r/xxfitness}) are positioned near the center,  bridging the muscular and thin ideals. The right end clusters forums focusing on the thin ideal (e.g., \texttt{r/1200isplenty}) and associated psychopathologies (e.g., \texttt{r/AnorexiaNervosa}).  Mental health subreddits (e.g., \texttt{r/SuicideWatch}) are nearby, reflecting shared membership with eating disorder communities.


\begin{figure}[ht]
  \centering
  \includegraphics[width=\columnwidth]{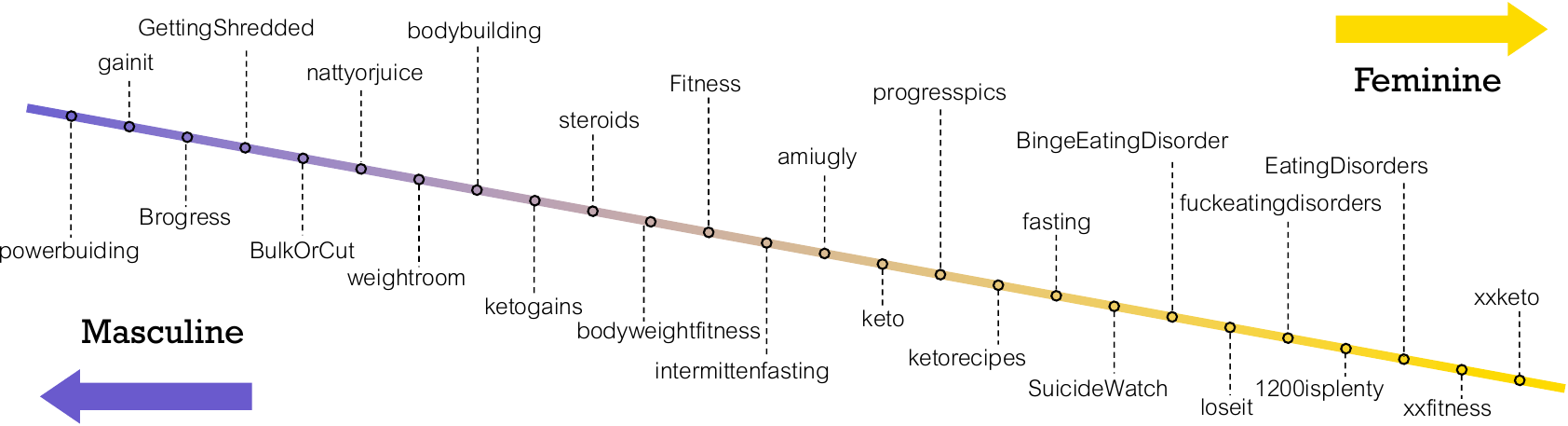}
  \caption{Ranking of our relevant communities along the masculine-feminine dimension.} 
  \label{fig:fem-masc-1d}
\end{figure}

\subsubsection{Gender Axis} 
This representation takes a binary approach to gender, spanning from masculine to feminine. While this conceptualization of gender is false, our goal here is to preliminarily explore how traditional gender roles intersect with body ideals at the community level, so we focus on that binary framing.

For the gender dimension, we reuse the axis introduced by \citet{waller2021quantifying}. Their method embeds Reddit communities in a user-subreddit graph and then defines a gender line using ten pairs of forums that are similar in topic but differ in the typical gender of participants (for example, \texttt{r/AskWomen} versus \texttt{r/AskMen}). A community’s gender score is its projection onto this line, so subreddits whose member base overlaps more with the ``women’s'' anchors fall on the feminine side, and those whose members overlap more with the ``men’s'' anchors fall on the masculine side.

Of the 46 subreddits in our corpus, 26 appear in \citet{waller2021quantifying} and can therefore be placed directly on this gender axis (Fig.~\ref{fig:fem-masc-1d}); the remaining 20 subreddits were not scored in their catalog and so do not have gender scores in our analyses. We list the overlapping subreddits in Appendix~\ref{app:social_dimension}.

For the 26 communities that have both gender scores and our body ideal scores, the two dimensions are strongly correlated (Pearson $r = 0.84$). As visible in Figs.~\ref{fig:thin-musc-1d}--\ref{fig:fem-masc-1d}, subreddits that explicitly endorse the thin ideal in their names (\texttt{r/EatingDisorders}, \texttt{r/loseit}, \texttt{r/fasting}) fall on the feminine side, while muscular ideal forums (\texttt{r/powerbuilding}, \texttt{r/Brogress}, \texttt{r/BulkOrCut}) fall on the masculine side. Notably, the most feminine-coded fitness subreddits in our sample (\texttt{r/xxketo}, \texttt{r/xxfitness}) do not sit at the thin ideal end of the body ideal axis, foreshadowing our later finding that some women-centered strength communities emotionally resemble masculine strength spaces.

\subsection{Emotion Analysis}
\label{sec:results_emotions_ED}
Emotions are the fundamental drivers of human interaction, even in online spaces. They influence how people communicate, respond to others, and perceive the overall community atmosphere. 
In this section, we identify and characterize differences in emotional expression along the body ideal spectrum. Further, we analyze engagement within communities by looking at the emotional tone of comments to highlight differences in emotional support.


\subsubsection{Differences in Emotional Expressions along the Body Ideal Spectrum}

Emotional expression varies significantly across subreddits. To highlight these differences, we exclude 25\% of the least emotional posts---submissions and comments in the top quartile of neutral emotion confidence scores. Fig.~\ref{fig:results_emotions_ED} in the Appendix shows the median confidence score of the neutral emotion (no emotion) for the remaining 75\% of submissions and comments in each subreddit, ranked along the body ideal spectrum. 
We observe a clear separation by body ideal: muscular-focused subreddits generally express less emotion than thin ideal forums discussing anorexia and other eating disorders. Mental health-related forums (\texttt{r/Suicidewatch}, \texttt{r/BDDVent}) exhibit the highest emotionality, reflecting their focus on deeply personal challenges.
Overall, comments are more neutral and less emotional than submissions, except in muscular ideal communities like \texttt{r/Brogress}, \texttt{r/weightroom}, and \texttt{r/progresspics}, where responses show heightened emotionality.

\begin{figure}[t]
  \centering
  \includegraphics[width=0.9\columnwidth]{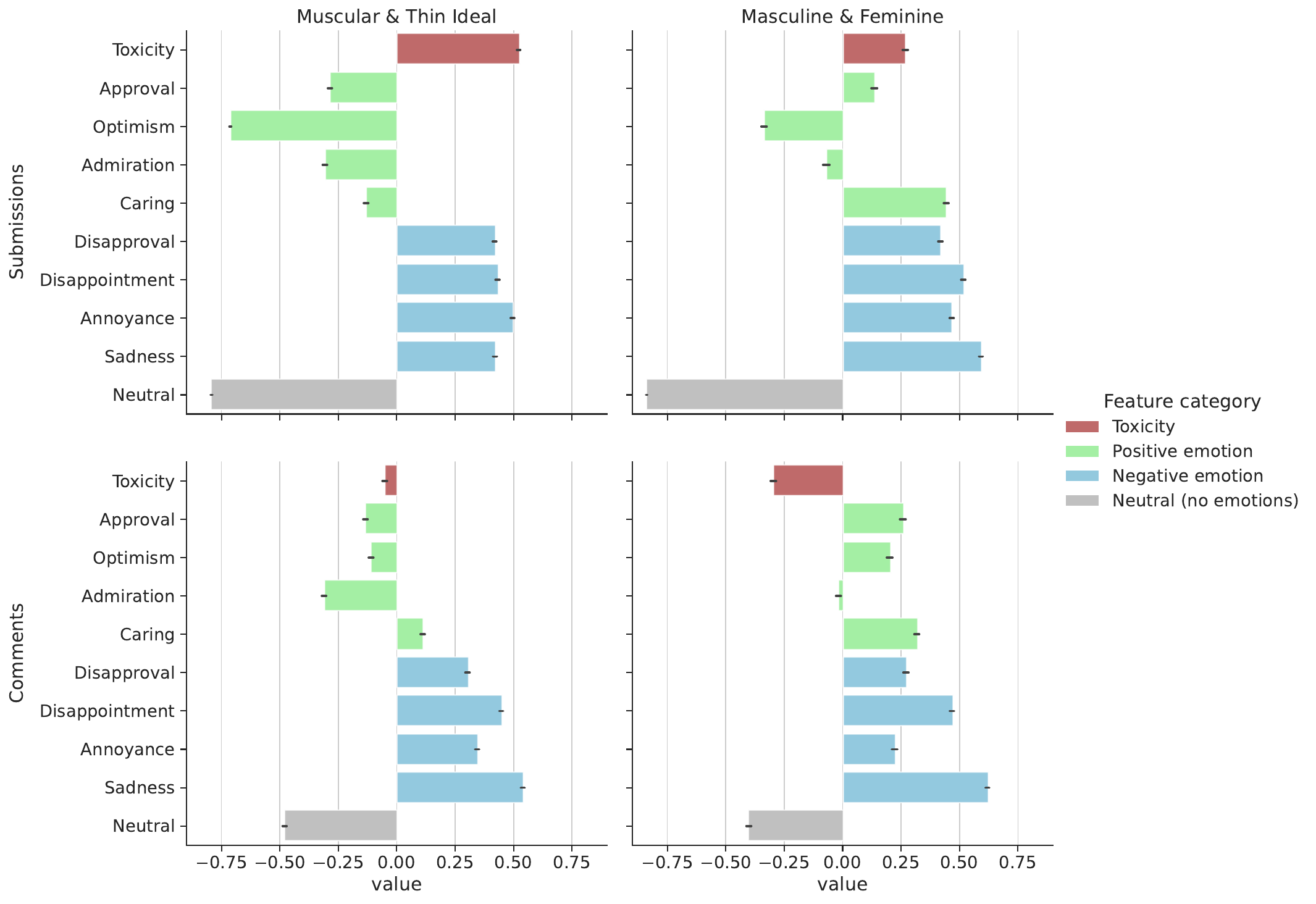}
  \caption{Spearman's correlation coefficient between (left) the body ideal scores and toxicity/emotion scores, and (right) the gender scores and toxicity/emotion scores of different communities in (top) submissions and (bottom) comments. 
  Emotions include the top 4 positive ones (approval, optimism, admiration, and caring) and the top 4 negative ones (disapproval, disappointment, annoyance, and sadness), with the highest median values in submissions and comments, and neutral.
  The analysis focuses on the top 75\% of data with the highest toxic and emotional content.
  Confidence intervals were obtained by 1000 bootstrap iterations.} 
  \label{fig:spearman_correlation}
\end{figure}

Fig.~\ref{fig:emotions_submissions_comments_filtered} in the Appendix shows average emotion confidence scores across subreddits ordered along the muscular-thin ideal spectrum. 
The heatmap reveals distinct emotional landscapes: ``thin ideal'' communities express more negative emotions (annoyance, disappointment, sadness) and toxic language, while ``muscular ideal'' forums favor positive emotions (approval, optimism, admiration, caring) or neutral language. Comments generally mirror the emotional patterns of submissions, though with more subtle distinctions along the body ideal spectrum. Notable differences in comments include a slight increase in caring expressions within thin ideal communities and higher toxicity in muscular ideal forums. We observe that the subreddits that are feminine but do not promote the thin ideal exhibit emotionality more in line with the muscular subreddits.

The ED and mental health communities share the language of psychological distress. To see this, we represent each subreddit as a vector of the emotion confidence scores (approval, admiration, joy, annoyance, disappointment, sadness) of the submissions. The TSNE \citep{van2008visualizing} visualization of the vector embeddings in Fig.~\ref{fig:emo_embeddings} reveals two distinct clusters: on the right is the ``mental health'' cluster consisting of eating disorder communities (\texttt{r/AnorexiaNervosa}, \texttt{r/bulimia}), suicidality, dysmorphia, and self-harm communities (\texttt{r/suicidewatch}, \texttt{r/MadeOfStyrofoam}, \texttt{r/BDDvent}).  In contrast, muscular ideal communities and non-ED thin ideal communities are separated from them in the embedding space. Despite promoting potentially problematic dietary and exercise regimens, they use distinctly different emotional language.

To quantify the variation in emotional expression across the body ideal spectrum, we show in Fig.~\ref{fig:spearman_correlation}(left) the Spearman correlation coefficient between median emotion confidence of submissions (top row) and comments (bottom row) and the body ideal spectrum scores. 
The ``thin ideal'' communities express more negative emotions (disapproval, disappointment, annoyance, sadness) and also use more toxic language, while the ``muscular ideal'' forums are more likely to express positive emotions (approval, optimism, admiration, caring) or no emotion at all (neutral). This pattern holds for both submissions and comments, suggesting that cross-community differences are not solely driven by some forums being used mainly for venting in posts and others for feedback in replies. Optimism is the positive emotion most strongly associated with the muscular ideal communities.

Emotional patterns in body image communities generally reflect broader gender differences in emotional expression~\citep{park2016women, brody2008gender}.  Fig. \ref{fig:spearman_correlation} (right) shows the correlation between subreddit emotions and their gender spectrum scores.
We observe that feminine-dominated communities exhibit a wider range of emotional expressions, while masculine-dominated forums display less emotionality overall. Optimism and admiration are the only emotions with stronger expression in masculine community submissions. We observe that the feminine subreddits promote the thin ideal (\texttt{r/xxketo}, \texttt{r/fitness}, and by extension \texttt{r/FlexinLesbians}) exhibit lower levels of emotion, just as muscular ideal communities.

\subsubsection{Emotions in Community Engagement along the Body Ideal Spectrum}
To explore differences in emotional support online communities provide for their members, we look at emotions expressed in submissions and comments separately.
The most common, or dominant, emotion in submissions is curiosity (Fig.~\ref{fig:dom_emo_thin-ideal}), although several different emotions appear to dominate across communities, from joy (e.g., \texttt{r/progresspics})  and amusement (e.g., \texttt{r/EDanonymemes}) to disappointment (e.g., \texttt{r/BDDvent}) and sadness (e.g., \texttt{r/suicidewatch}). 
The emotional tone of comments is similar to that of submissions (Fig.~\ref{fig:spearman_correlation}), although more subdued. Fewer emotions are expressed in the comments compared to submissions. The dominant emotion in comments is limited to admiration, approval, and caring. The dominant emotion of comments at the ``thin ideal'' end of the body ideal spectrum is caring, suggesting that these communities provide emotional support for their peers. In contrast, the dominant emotion of comments on the ``muscular ideal'' end is neutral or admiration, with more than half of the comments in \texttt{r/progresspics} and \texttt{r/Brogress} expressing admiration, reflecting the importance of peer approval in hypermuscular  communities~\citep{frederick2017precarious}. This behavior is also exhibited by the non-thin ideal feminine subreddits.

\subsection{Toxicity Analysis}
Toxic language has been associated such as harassment, trolling, and discriminatory language online~\citep{sheth2022defining, he2023cpl}. Unchecked toxicity can rapidly erode the quality of online interactions~\citep{chang2023feedback} and harm mental health~\citep{chen2023uncivil,kast2018unspoken}.
Toxicity is often used as a proxy of harm~\citep{ferra2021, 10.1145/3603399, 10.1145/3422841.3423534}, although more recent studies caution that this pattern fails in online spaces serving marginalized identities~\citep{dorn2023non}.

\subsubsection{Toxicity and Online Harms}
\begin{figure}[t]
  \centering
  {}
  \includegraphics[width=0.8\columnwidth]{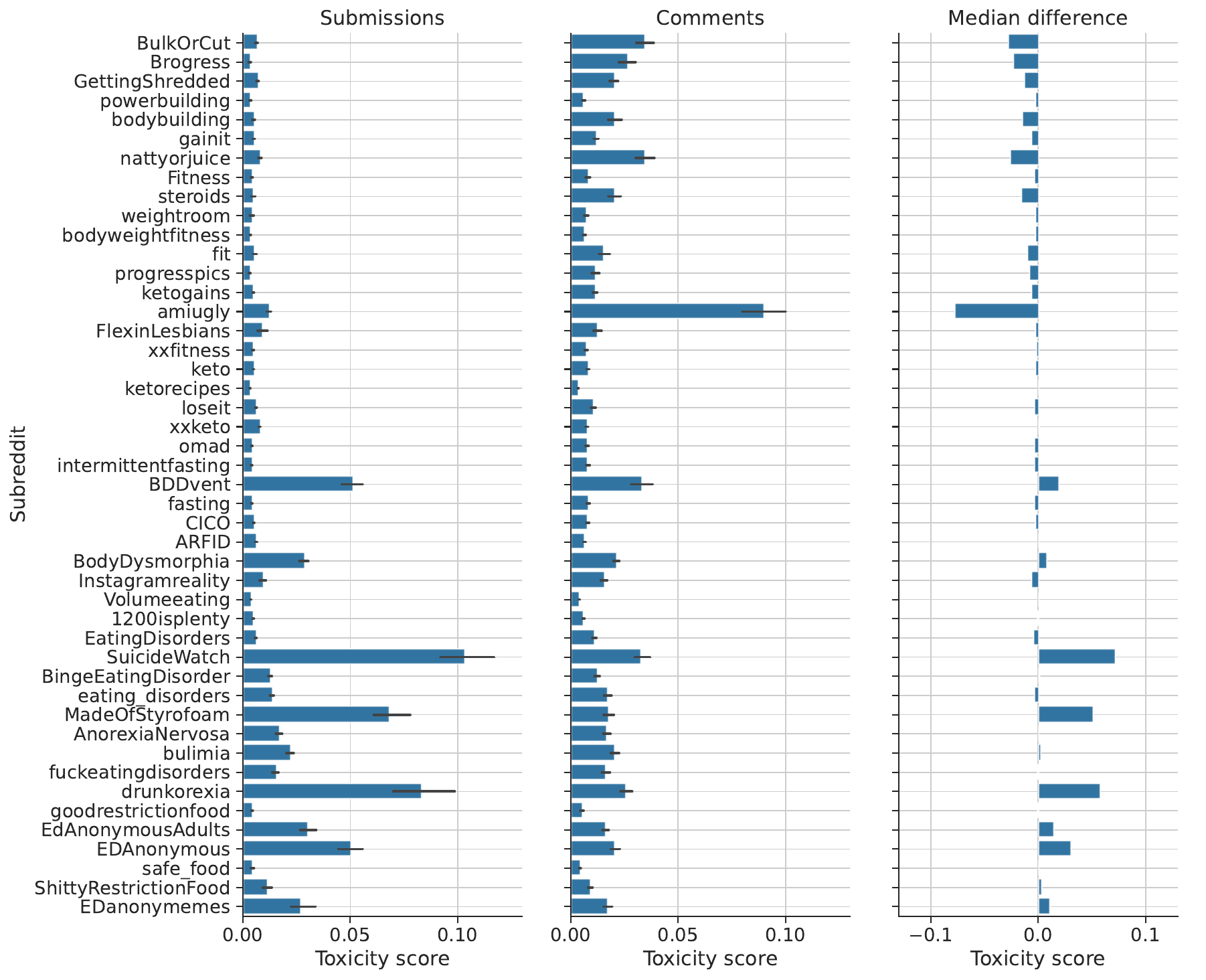}
  \caption{Distribution of toxicity scores in subreddits, ordered according to the muscular-thin ideal dimension. The bars show the median confidence values of toxicity in submissions (left), comments (middle), and their difference (right), in different subreddits. The analysis focuses on the top 75\% of data with the highest toxic content, excluding submissions and comments with a toxicity score below the 25th percentile separately. } 
  \label{fig:results_toxicity_ED_gender}
\end{figure}

We calculate toxicity scores for submissions and comments. Fig.~\ref{fig:results_toxicity_ED_gender} shows toxicity scores across all subreddits in our data, ordered by the body ideal, highlighting differences in their median scores. Note, we include scores of the top 75\% most toxic submissions and comments.
As shown in Fig.~\ref{fig:results_toxicity_ED_gender}, forums dedicated to mental health and eating disorders exhibit high levels of toxicity in both submissions and comments, whereas goal-oriented strength and diet forums such as \texttt{r/ketogains} and appearance-sharing forums such as \texttt{r/progresspics} show similar toxicity profiles despite serving different conversational functions. Closer examination reveals that, despite being flagged as highly toxic, these subreddits are heavily moderated to maintain a safe environment without spreading harm. For example, \texttt{r/SuicideWatch} provides a safe space for peer support for individuals struggling with suicidal thoughts. To minimize harm, moderators enforce strict rules to ensure safety and civility, such as avoiding inciting language and judgmental comments. Users in these forums often seek emotional support through self-disclosure, which may include explicit language (e.g., obscenities). Toxicity detection algorithms are prone to incorrectly flag such texts as toxic \citep{garg2023handling}. Therefore, such posts are frequently flagged as toxic. Below is an original post from \texttt{r/SuicideWatch} illustrating this point:

\begin{quote}
\small
\texttt{r/SuicideWatch}: \textit{``\textbf{Fuck everyone who has hurt me} \\ Fuck you all. Fuck you piece of shit medical legal assholes. I hope you all live a long and ugly life full of suffering. I hope you go to bed every night crying because of the evil you emit. Fuck my ``family''. Fuck all you former friends who left when it got hard. Fuck ``doctors'', ``judges'', and whatever other little bitches out there who are just stuck up on that high horse of theirs. FUCK YOU ALL, GO KILL YOURSELVES. But do it away from me so I can die in peace.''}
\end{quote}

Conversely, forums focusing on the muscular ideal, such as \texttt{r/steroids} and \texttt{r/GettingShredded}, present a higher risk of harm despite having lower toxicity scores compared to other communities. 
Specifically, \texttt{r/steroids} features discussions promoting the use of illegal muscle-building substances with significant health risks \citep{parssinen2002steroid}. The (mis)information and advice shared on these forums often contradict medical guidelines, posing serious physical harm to participating users.

\begin{quote}
\small
\texttt{r/steroids}: \textit{``\textbf{[Compounds] Experience threads for common stacks} \\ I think a great idea for new compound experience threads could be really common stacks like nandrolone/dbol or tren/mast, etc. It could give people good insight into what the compounds can do differently when combined with other complementary compounds; it could help inform people of the specific synergies present between certain compounds. Just an idea.''}
\end{quote}

\label{sec:results_toxicity_ED}

\subsubsection{Toxicity in Community Engagement}
As shown in Fig.~\ref{fig:results_toxicity_ED_gender}, there is little difference in toxicity scores between submissions and comments. However, significant deviations from this trend exist. Specifically, members of the mental health and ED communities typically respond with less toxicity. This is likely because submissions in these forums often express distress, prompting supportive responses. In contrast, comments in muscular-focused communities like \texttt{r/steroids}, \texttt{r/Brogress}, \texttt{r/BulkOrCut}, and \texttt{r/bodybuilding} tend to be more toxic than submissions. This trend is also evident in appearance-focused subreddits such as \texttt{r/amiugly} and \texttt{r/progresspics}, where users invite peers to comment on their appearance. Interestingly, comments in these communities often exhibit a mix of toxicity and admiration.
In muscular ideal forums, individuals' body self-disclosures receive validation, often using explicit language that our toxicity model flags as highly toxic. 
The following examples illustrate cases where profanity is part of coarse camaraderie and amplifies admiration and support rather than hostility:
\begin{quote}
\small
\texttt{r/progresspics}: \textit{``You look fucking phenomenal''} \\
\texttt{r/Brogress}: \textit{``Fucking amazing bro. Keep it up. Looking big man''} \\
\texttt{r/amiugly}: \textit{``You’re so handsome holy shit''}
\end{quote}
To an automated classifier, these strings are ``toxic,'' but in gym and ``bro'' subcultures, they function as affiliative praise. This nuance helps explain the reversal in the association of toxicity with gender and body ideal when comparing submissions with comments (Fig.~\ref{fig:spearman_correlation}).

\begin{figure}[t]
  \centering
  \includegraphics[width=0.75\columnwidth]{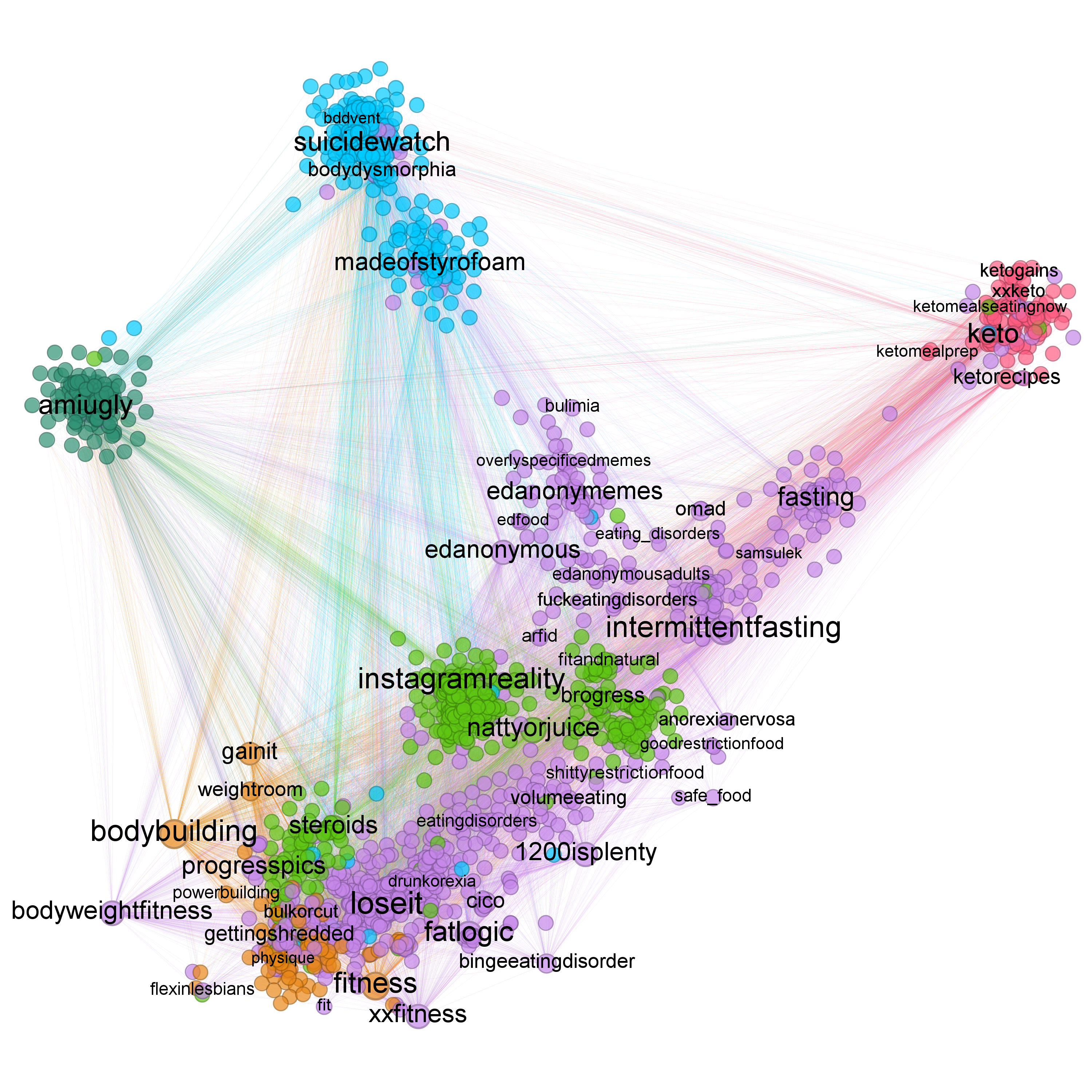}
  {}
  \caption{Network of subreddit mentions. 
  Each point is a subreddit, with edges linking sources to their mentioned subreddits.
  Node colors represent higher-level clusters, with link colors matching the source subreddit. Node sizes reflect their degrees. The light blue cluster focuses on mental health, dark pink on the keto diet, dark green on body image concerns, purple on extreme diets and eating disorders, and orange and light green on bodybuilding, fitness, and physique goals.
  } 
  \label{fig:results_mention_network}
\end{figure}

\subsection{Structural Analysis}
To understand how body-ideal communities are embedded within the broader Reddit ecosystem, we examine two structural dimensions: the subreddit mention network, which reveals referral and support pathways, and text-based semantic similarity, which identifies clusters of shared discourse. 
Together, our results reveal that gendered norms not only shape how distress is linguistically expressed but also manifest in the organizational structure of online communities, influencing users’ access to mental health support. 

\subsubsection{Subreddit Mentions Network}
\label{sec:results_mentions_network}
We examine the emergent organization of online communities, as revealed via the subreddit mention network (see Methods). The mention network provides a structural view of peer support and how online communities not only exchange information but also direct users toward (or away from) spaces where psychological distress is acknowledged, and mental health support is available.

Using regex matching, we extracted 18,202 subreddit mentions from our 54 focal subreddits, 1950 of which are unique. We remove subreddits mentioned fewer than 10 times. Of the mentioned subreddits, 71 have been banned ({removed for severe policy violations, inaccessible to all users}), 18 were gated ({access restricted only to approved users}), and three were quarantined ({hidden from search and flagged for harmful content}). This network expands our original set of subreddits by surfacing frequently mentioned forums and highlights attention and support pathways: some communities mainly direct readers outward, while others are destinations that many communities point toward. In manual checks, mentions included both housekeeping messages (for example, ``this post is better suited for \texttt{r/loseit}'') and referrals to support forums; removing common housekeeping phrases did not change the cluster structure. Fig.~\ref{fig:results_mention_network} plots this network, with only the original 54 subreddits labeled.

Community detection in the mention network revealed marked gendered asymmetries in access to support. Subreddits associated with thinness-oriented body ideals---and hence more feminine membership---such as those focused on restrictive dieting, weight loss, or anorexia, were tightly connected to mental health cluster (light blue) of forums addressing suicide (\texttt{r/SuicideWatch}), self-harm (\texttt{r/MadeOfStyrofoam}), and body dysmorphia (\texttt{r/BodyDysmorphia}, \texttt{r/BDDVent}). These links suggest that users in thinness-oriented communities not only express greater emotional distress linguistically but are also structurally closer to support pathways, consistent with gendered norms of expressiveness and help-seeking.

In contrast, muscularity-oriented forums, including more masculine bodybuilding and bigorexia spaces, were embedded in dense clusters with few links to mental health subreddits, despite promoting behaviors associated with body dysmorphia and psychological impairment. As a result, individuals experiencing muscularity-related distress may be less visible to peer support networks and less likely to be directed toward mental health resources.


\begin{figure}[ht]
  \centering
  \includegraphics[width=0.8\columnwidth]{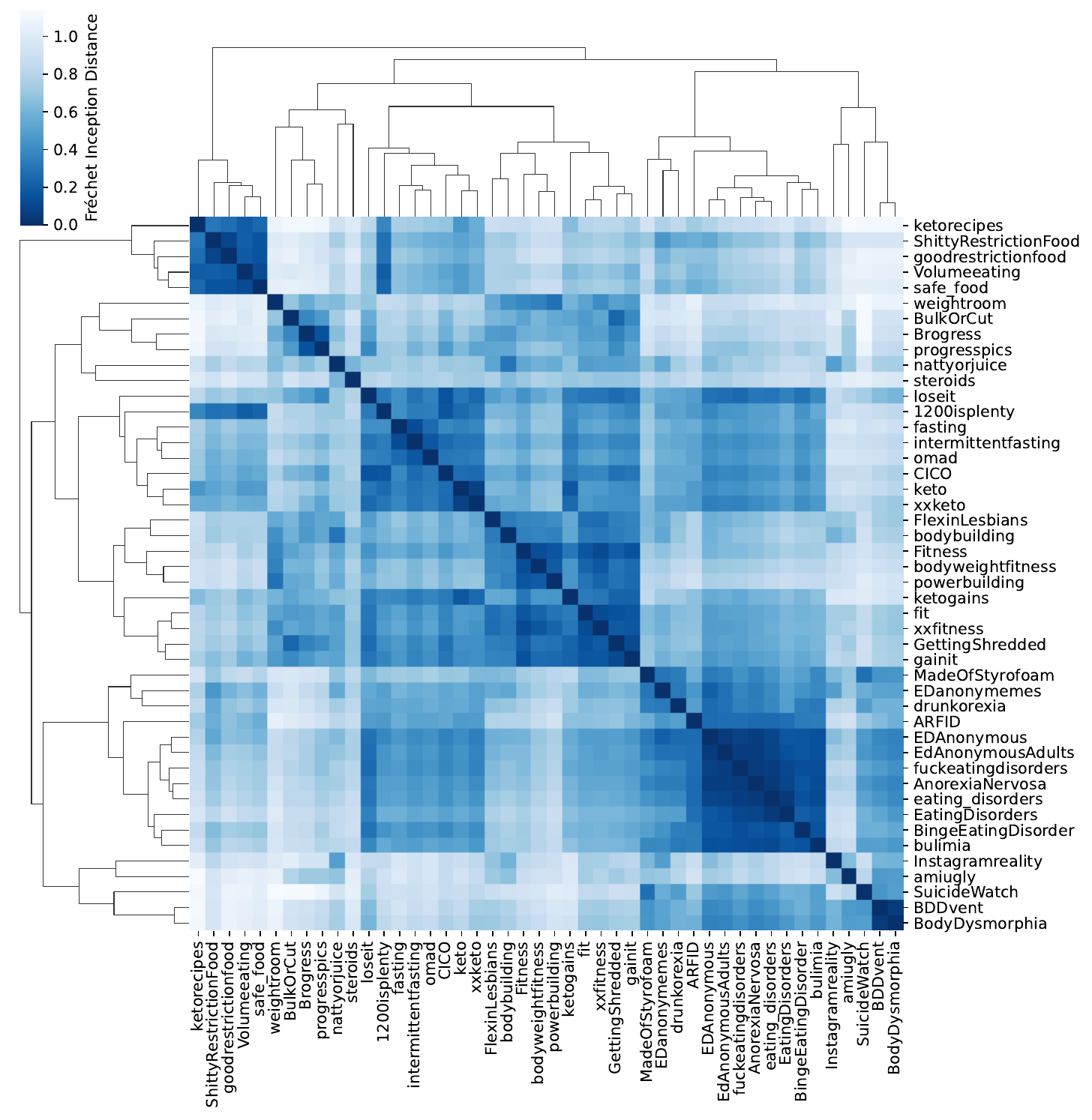}
  {}
  \caption{Hierarchical agglomerative clustering of diet and fitness subreddits by text similarity in submissions, based on the All MPNet Base V2 sentence embedding model \protect\citep{10.5555/3495724.3497138} and the Fréchet Inception Distance \citep{10.5555/3295222.3295408,kour-etal-2022-measuring}. Cells with dark colors indicate more similarity (less distance) between the subreddits. 
  } 
  \label{fig:text_similarity}
\end{figure}

\subsubsection{Semantic Similarity} \label{sec:results_text_similarity}

To further examine how gendered norms shape the visibility of psychological distress across body-ideal communities, we analyzed the semantic similarity of subreddit content (Fig.~\ref{fig:text_similarity}). The heatmap reveals five coherent clusters of semantically similar content, including restrictive dieting communities, a large mixed cluster of dieting and fitness subreddits, and a distinct cluster (bottom-right) containing eating disorder and mental health forums. Notably, muscularity-oriented communities cluster with fitness and performance-based groups rather than with ED or mental health communities, despite evidence that users in these spaces engage in behaviors consistent with body dysmorphia. Their language had little semantic overlap with discussions of distress, impairment, or help-seeking.

This semantic separation provides a complementary perspective to our emotional tone analysis: the absence of psychological distress in muscularity-oriented communities is not only a matter of reduced emotional expressiveness but also reflects the substantive topics around which these communities organize. Whereas thinness-oriented communities share considerable semantic space with mental health subreddits, indicating overlapping concerns and community membership around body image, dysmorphia, and emotional vulnerability, muscularity-oriented communities remain linguistically and topically insulated from such discussions. This structural and semantic distancing may reinforce gendered norms that discourage acknowledgment of distress in masculine-leaning spaces and may hinder recognition of muscle dysmorphia as a mental health concern within these communities.

\subsection{Discussion}
Online diet and fitness communities reflect and amplify gendered sociocultural body ideals. By examining 46 Reddit communities, we found that thinness-oriented spaces were predominantly feminine-leaning, whereas muscularity-oriented communities had more masculine membership. These alignments echo broader cultural norms linking femininity to thinness and emotional expressiveness, and masculinity to muscularity, self-control, and emotional restraint \citep{Calogero2010,murnen2012body}.

Thinness-oriented communities, including those focused on restrictive dieting, anorexia, and bulimia, clustered closely with mental health and body dysmorphia forums in both the subreddit mention network and semantic similarity analyses. Their discourse was marked by high emotional expressiveness, particularly sadness and disappointment, and peers frequently responded with validation and referrals to support resources. These patterns are consistent with feminine socialization and with prior findings that women more readily recognize and disclose body image distress \citep{Graham_caring,chancellor2016recovery}.

Muscularity-oriented communities displayed a very different profile. They expressed far less emotion overall, and when emotions did appear, they tended toward admiration, encouragement, or banter. Linguistically and structurally, these communities were largely isolated from mental health spaces despite discussing behaviors consistent with muscle dysmorphia. Both the mention network and semantic similarity analyses showed that muscularity communities clustered with performance- and discipline-oriented forums rather than with eating disorders or support communities. These findings parallel research showing that traditional masculine norms discourage emotional expression and help-seeking, contribute to underdiagnosis of muscularity-oriented disorders, and stigmatize acknowledging distress \citep{Raisanen2014Role,Strother2012Eating,Seidler2016Role}.


Our findings show that online communities reproduce gendered norms surrounding bodies, emotions, and help-seeking. Thinness-oriented spaces foster visibility of psychological distress and connections to mental health support, whereas muscularity-oriented spaces obscure distress through emotional restraint, topical focus, and limited structural links to support communities. Online platforms offer a rich source of naturalistic behavioral data, capturing how body image concerns, coping strategies, and peer dynamics unfold in real time and at scale. Leveraging these digital traces can advance theoretical models of body image and eating disorders by revealing patterns of internalization, comparison, and support-seeking that are difficult to observe through traditional methods.

\section{Limitations and Ethical Considerations}

\paragraph{IRB Review and Privacy}  
This research touches on highly sensitive topics related to mental health, calling for extra precautions to minimize risks to study subjects as well as researchers.
To minimize privacy risks, identifiable information was removed, and analysis was carried out on aggregated data. 
The study protocol was reviewed by the authors' IRB. All data used for this study is public and collected following Reddit's terms of service.

\paragraph{Sex and Gender Considerations}
This study analyzes publicly available Reddit data where user demographics, including sex and gender, are not systematically collected or verified. Our gender dimension analysis utilizes computational methods to map communities along a masculine-feminine spectrum based on user co-activity patterns \citep{waller2021quantifying}, examining the intersection of traditional Western gender roles, body ideals, and online behavior. We acknowledge important limitations: the binary conceptualization of gender does not capture the full spectrum of gender identities, individual users' actual gender identities remain unknown, and dominant representation in media reflects cisgender and heteronormative cultural standards that may not adequately capture experiences of gender-diverse populations or intersectional identities. Future research should incorporate more inclusive approaches to understanding how diverse gender identities intersect with body image concerns in online spaces and center systems that are typically understudied.

\paragraph{Selection of subreddits and seed pairs}
This work relies on the semi-manual identification of relevant online forums. While we tried to compensate for the ad hoc nature of manual selection by systematically identifying additional forums using the subreddit mention network, the list of subreddits may still be incomplete, limiting our conclusions.

\citet{waller2021quantifying} identify exemplar communities for the gender dimension and algorithmically augment it with additional similar pairs of communities, to ensure that the dimension is not overly tied to idiosyncrasies of the manually selected seed communities.
However, we manually select the four pairs for the body ideal dimension without augmenting them, because the number of subreddits in our study (46) is relatively small compared to \citet{waller2021quantifying}'s (10,006), making the augmentation algorithm impractical. Furthermore, we carefully refined our selection of seed pairs through multiple revision rounds in consultation with various advisors. Table \ref{tab:seed_pairs} in the Appendix describes our rationale and the description of the seed communities.

\paragraph{Algorithmic Bias in Classifiers}
Emotion and toxicity classifiers contain biases in their training data and results \citep{he2024whose}. Thus, any emotion and toxicity system used will likely vary systematically across gender and racial dimensions.
However, our results show distinct communities where muscular ideal forums with feminine membership look more like other muscular forums rather than the thin ideal forums, implying robustness. 
We look forward to future work making these classifiers perform well across demographic variance. Our emotion findings should therefore be read as characterizing how different communities talk about emotion, not as claims about underlying physiological or felt emotional intensity.

\section{Conclusion}

Our analysis of Reddit's diet and fitness communities reveals a spectrum of body image concerns, from the thin ideal to the muscular ideal. We identified a gender-based emotional divide: thinness-oriented subreddits, often aligned with feminine communities, exhibit higher negative affect and toxicity compared to muscular ideal spaces. This pattern extends to mental health-focused communities, which show similarities to thin ideal spaces.

Online platforms provide an unparalleled window into how body image concerns, coping strategies, and peer dynamics unfold in everyday settings, offering rich behavioral data that can refine and expand existing theories of body image and eating disorders. By capturing real-time expressions of distress, support, and identity across diverse communities, these data can reveal mechanisms that traditional clinical or survey methods may miss. Harnessing such insights can inform the development of more effective moderation policies. Future work should explore the role of intersectional identities—such as race, gender, and socioeconomic status—in shaping emotional expression and support within these communities, analyze spaces serving marginalized groups, and develop interventions to counteract gender stereotypes and support diverse body image concerns.

\section*{Data Availability Statement}
The data supporting this study are publicly available Reddit data collected via Academic Torrents using the Pushshift API. Due to the sensitive nature of mental health discussions in the analyzed content, processed and relevant code will be made available to qualified researchers upon reasonable request for research purposes. Requests should be directed to the corresponding author and will be evaluated considering ethical guidelines for research involving vulnerable populations.




\bibliographystyle{apacite}
\bibliography{references,lerman}

\appendix

\section{Data Collection}

\begin{table}[ht]
\small
\begin{tabular}{p{0.15\linewidth}p{0.8\linewidth}}
\toprule
\textbf{Category} & \textbf{Keywords} \\ \hline
\textbf{Muscularity} & \emph{shredded}, \emph{swole}, \emph{bodybuilding}, \emph{mensphysique}, \emph{aggressivecut}, \emph{aggressivebulk}, \emph{macrotracking}, \emph{shreddeddiet}, \emph{trainuntilfailure}, \emph{pushyourlimit}, \emph{norestday}, \emph{anabolicstack}, \emph{steroids}, \emph{tren}, \emph{anabolicgear}, \emph{proteinpowder}, \emph{whey}, \emph{creatine}, \emph{preworkout}, \emph{cheatmeal}, \emph{cheatday}, \emph{gymmotivation}, \emph{gymmasculinity}, \emph{alphamalegym}, \emph{beaman}, \emph{embracemasculinity}, \emph{workoutbreakup}, \emph{gymtherapy}, \emph{liverkingdiet}, \emph{rawmeatdiet}, \emph{TRT}, \emph{testosterone}, \emph{hormoneboost}, \emph{musclesdissatisfaction}, \emph{muscledysmorphia}, \emph{badmen}. \\ \hline
\textbf{Thinness} & \emph{thinspo}, \emph{edtwt}, \emph{proana}, \emph{proanatips}, \emph{anatips}, \emph{meanspo}, \emph{fearfood}, \emph{sweetspo}, \emph{eatingdisorder}, \emph{bonespo}, \emph{promia}, \emph{redbracetpro}, \emph{m34nspo}, \emph{fatspo}, \emph{lowcalrestriction}, \emph{edvent}, \emph{WhatIEatInADay}, \emph{Iwillbeskinny}, \emph{thinspoa}, \emph{ketodiet}, \emph{skinnycheck}, \emph{thighgapworkout}, \emph{bodyimage}, \emph{bodygoals}, \emph{weightloss}, \emph{skinnydiet}, \emph{chloetingchallange}, \emph{fatacceptance}, \emph{midriff}, \emph{foodistheenemy}, \emph{cleanvegan}, \emph{keto}, \emph{cleaneating}, \emph{intermittentfasting}, \emph{juicecleanse}, \emph{watercleanse}, \emph{EDrecovery}, \emph{bodypositivity}, \emph{dietculture}, \emph{slimmingworld}, \emph{losingweight}, \emph{weightlossmotivation}, \emph{healthyliving}, \emph{weightlosstips}, \emph{weightlossjourney}, \emph{wegovy}, \emph{semaglutide}, \emph{ozempic}. \\ \bottomrule
\end{tabular}
\caption{Search terms used to retrieve body image disorder content from social media, separated by muscularity-focused and thinness-focused keywords.}
\label{tab:search-terms-separated}
\end{table}

\section{Analysis of Subreddits}



\subsection{Seed Pairs of Subreddits}

Table~\ref{tab:seed_pairs} shows the seed pairs used to construct the body ideal spectrum.

\begin{table}[h]
\small
\begin{tabular}{|p{0.3\linewidth}|p{0.6\linewidth}|}
\hline
\textbf{Seed Pairs}                   & \textbf{Description}                                                                                                                                                                                               \\ \hline
\texttt{r/loseit}, \texttt{r/gainit}                    & Discussing healthy methods and members' progress of losing/gaining weight.                                                                                                                                         \\ \hline
\texttt{r/intermittentfasting}, \texttt{r/steroids}   & Extreme measures (long periods of fasting/using anabolic supplements) to lose weight or gain muscles.                                                                                                              \\ \hline
\texttt{r/AnorexiaNervosa}, \texttt{r/GettingShredded}& Discussions around unhealthy obsessions with the thin/muscular ideals.                                                                                                                                             \\ \hline
\texttt{r/Instagramreality}, \texttt{r/nattyorjuice}    & Fact-checking and discussing people's body images online: whether they use photoshop their pictures to look skinnier (\texttt{r/InstagramReality}) or whether they are on anabolic supplements to look muscular (\texttt{r/nattyorjuice}). \\ \hline
\end{tabular}
\caption{Description of seed pairs of subreddits to construct the body ideal spectrum. }
\label{tab:seed_pairs}
\end{table}

\begin{figure}[tbh]
  \centering
  \includegraphics[width=0.8\columnwidth]{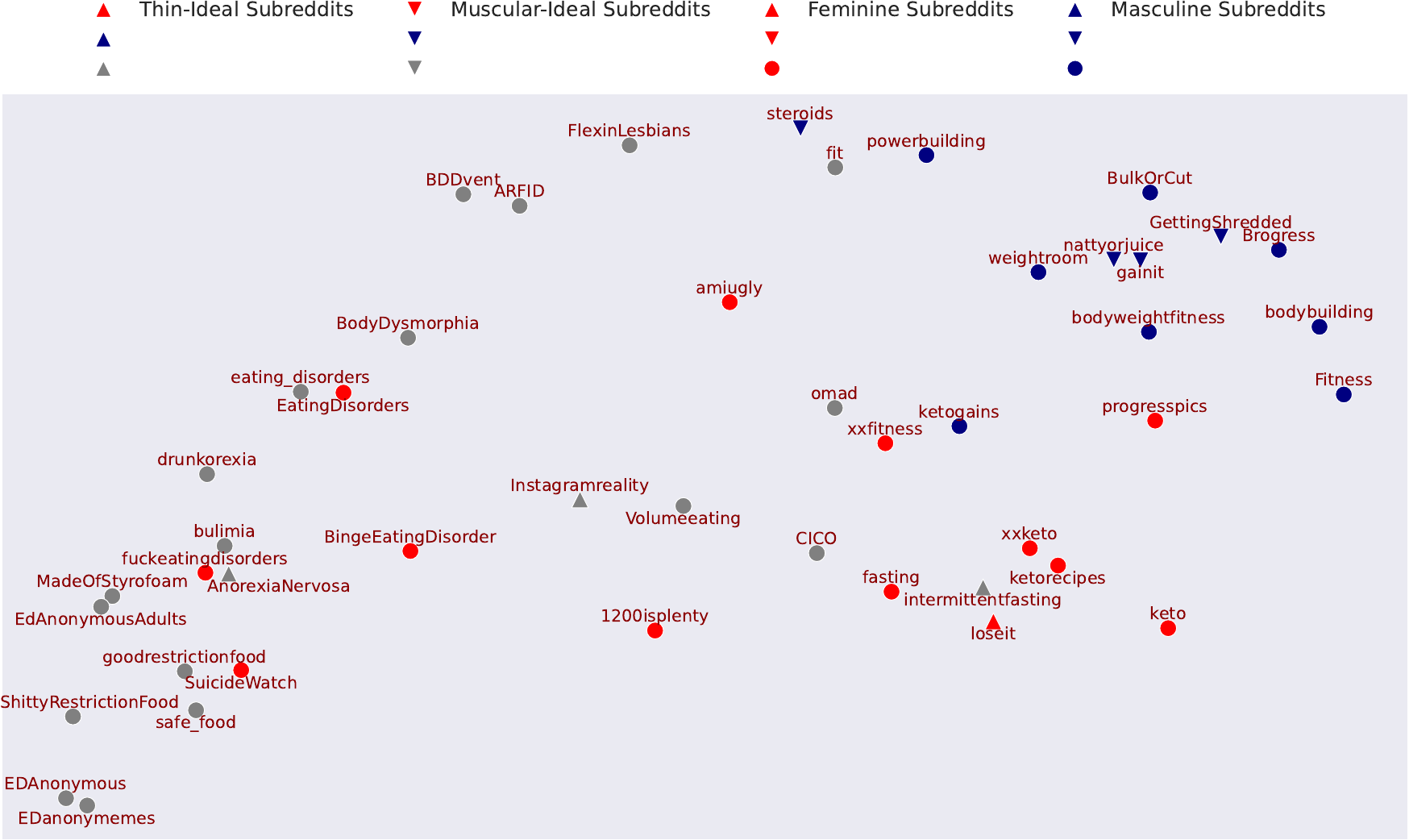}
  
  \caption{TSNE embeddings of communities from the user co-activity network. 
  For the body ideal spectrum, the identified thin/muscular-ideal communities in the seed pairs to construct the axis are marked by upper/lower triangles, and the remaining communities are marked by circles.
  For the gender spectrum, the feminine/masculine communities are colored in red/blue, and the remaining communities not studied by \citet{waller2021quantifying} are colored in gray. } 
  \label{fig:tsne_emb_social_dim}
\end{figure}

\subsection{Subreddits along the Dimensions of Gender and Body Ideal}
\label{app:social_dimension}
TSNE embeddings of communities from the bipartite user co-activity (posting and commenting) network are shown in Fig. \ref{fig:tsne_emb_social_dim}. The distribution of communities along the masculine-feminine dimension is shown in Fig. \ref{fig:fem-masc-1d}.

\begin{figure}[tbh]
  \centering
\includegraphics[width=0.8\columnwidth]{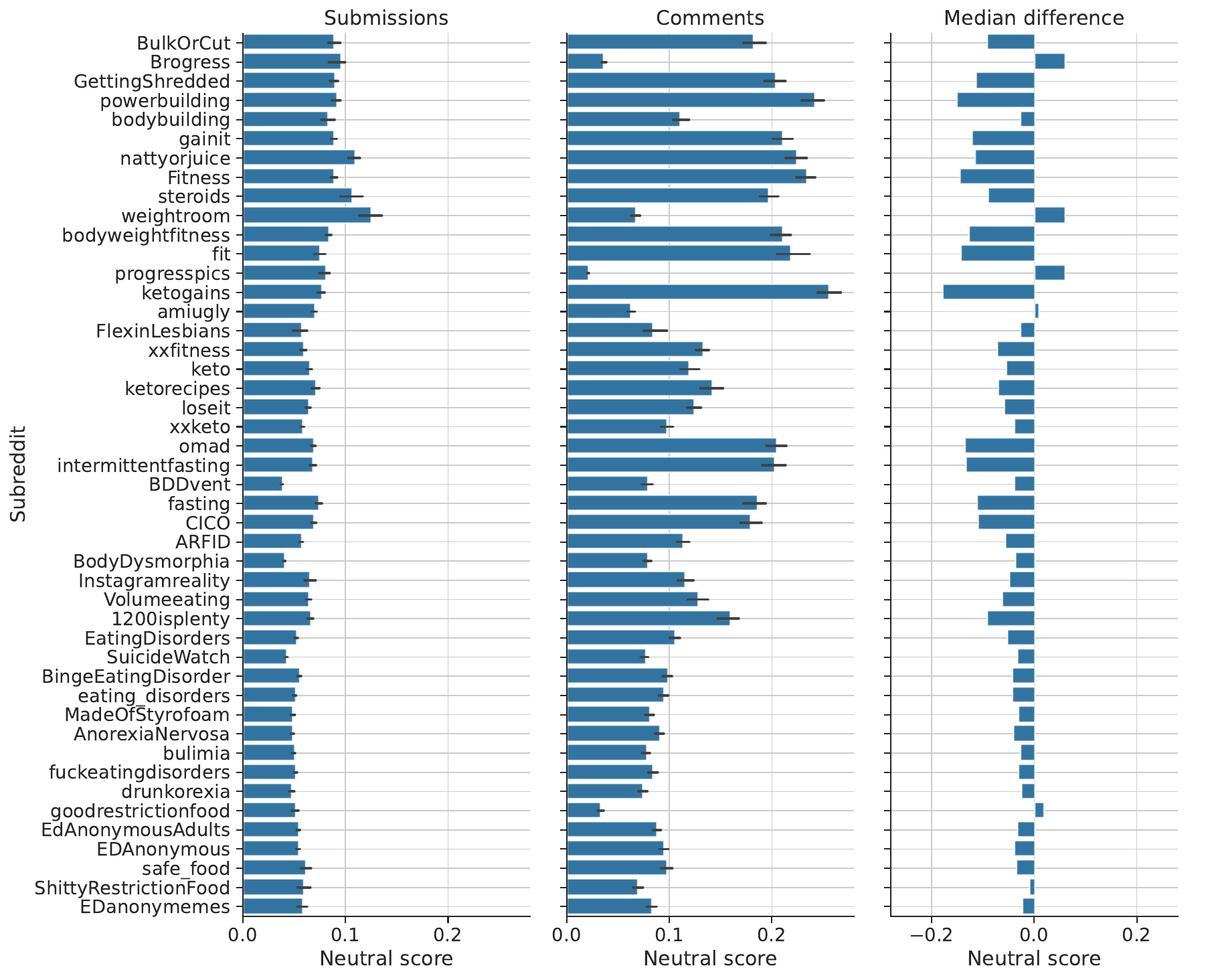}
  
  \caption{Distribution of neutral emotion in subreddits, ordered according to the muscular-thin ideal dimension. The bars show the median confidence values of the neutral emotion (i.e., that the post expresses no emotion) in submissions (left), comments (middle), and their difference (right), in different subreddits. Larger values indicate less emotionality. The analysis focuses on the top 75\% of data with the highest emotional content, excluding submissions and comments with a neutral score above the 75th percentile separately. 
  } 
  \label{fig:results_emotions_ED}
\end{figure}

\subsection{Emotional Landscape of the Subreddits}
\label{app:emo_subreddits}

\begin{figure}[ht] 
  \centering
  \includegraphics[width=\columnwidth]{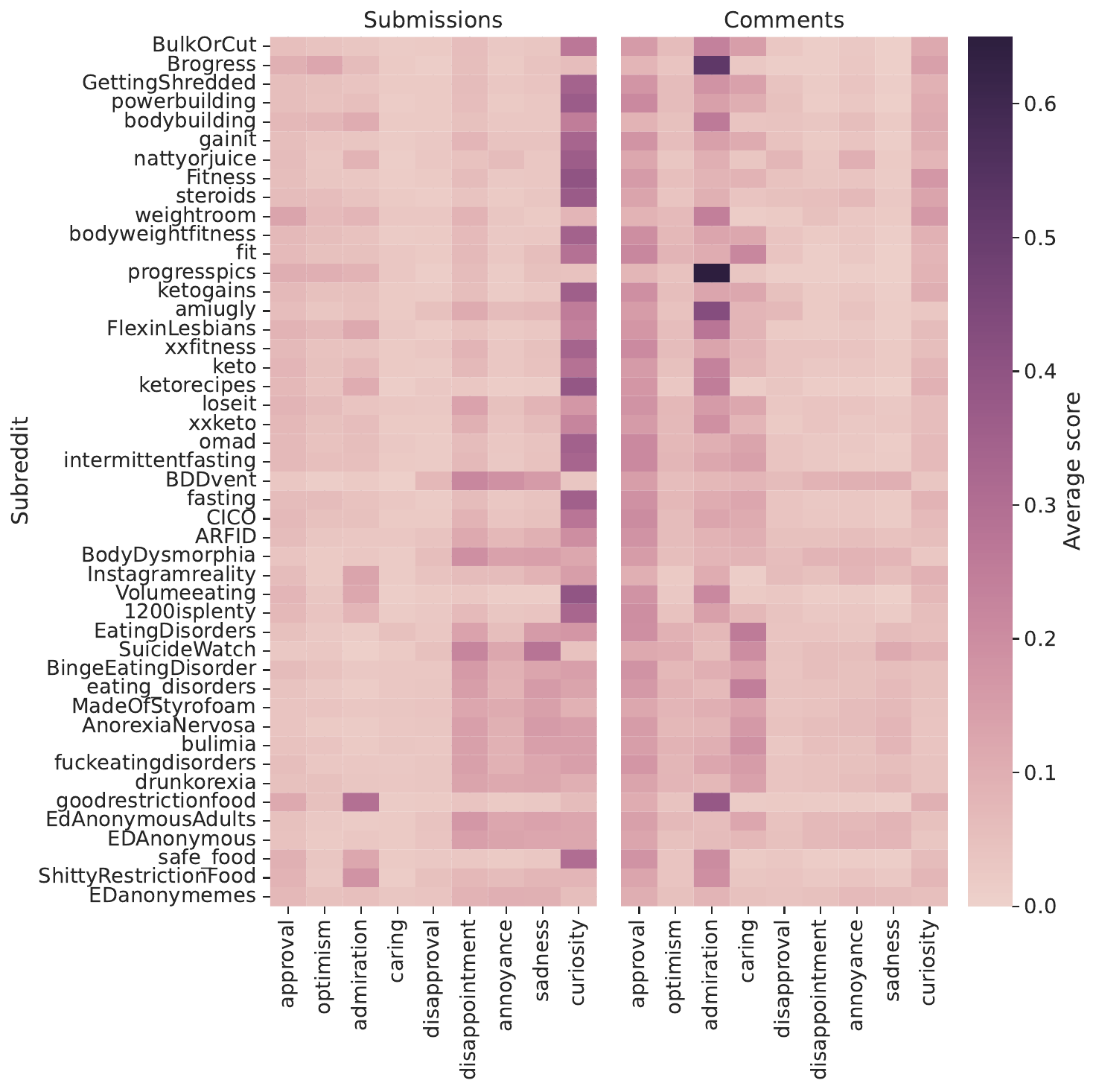}
  
  \caption{Average score of the top positive emotions (approval, optimism, admiration, and caring), the top negative emotions (disapproval, disappointment, annoyance, and sadness), and the ambiguous emotion curiosity, in submissions (left) and comments (right) in different subreddits ordered according to the muscular-thin ideal dimension. This average reflects the average score of the filtered set: top 75\% of data with the highest emotional content. 
  } 
  \label{fig:emotions_submissions_comments_filtered}
\end{figure}

Fig. \ref{fig:emotions_submissions_comments_filtered} shows the average score of positive emotions (approval, optimism, admiration, and caring), negative emotions (disapproval, disappointment, annoyance, and sadness), and the ambiguous emotion of curiosity, in submissions and comments in subreddits ordered according to the
muscular-thin ideal dimension.


%

\begin{figure}[t]
  \centering
  \includegraphics[width=0.94\columnwidth]{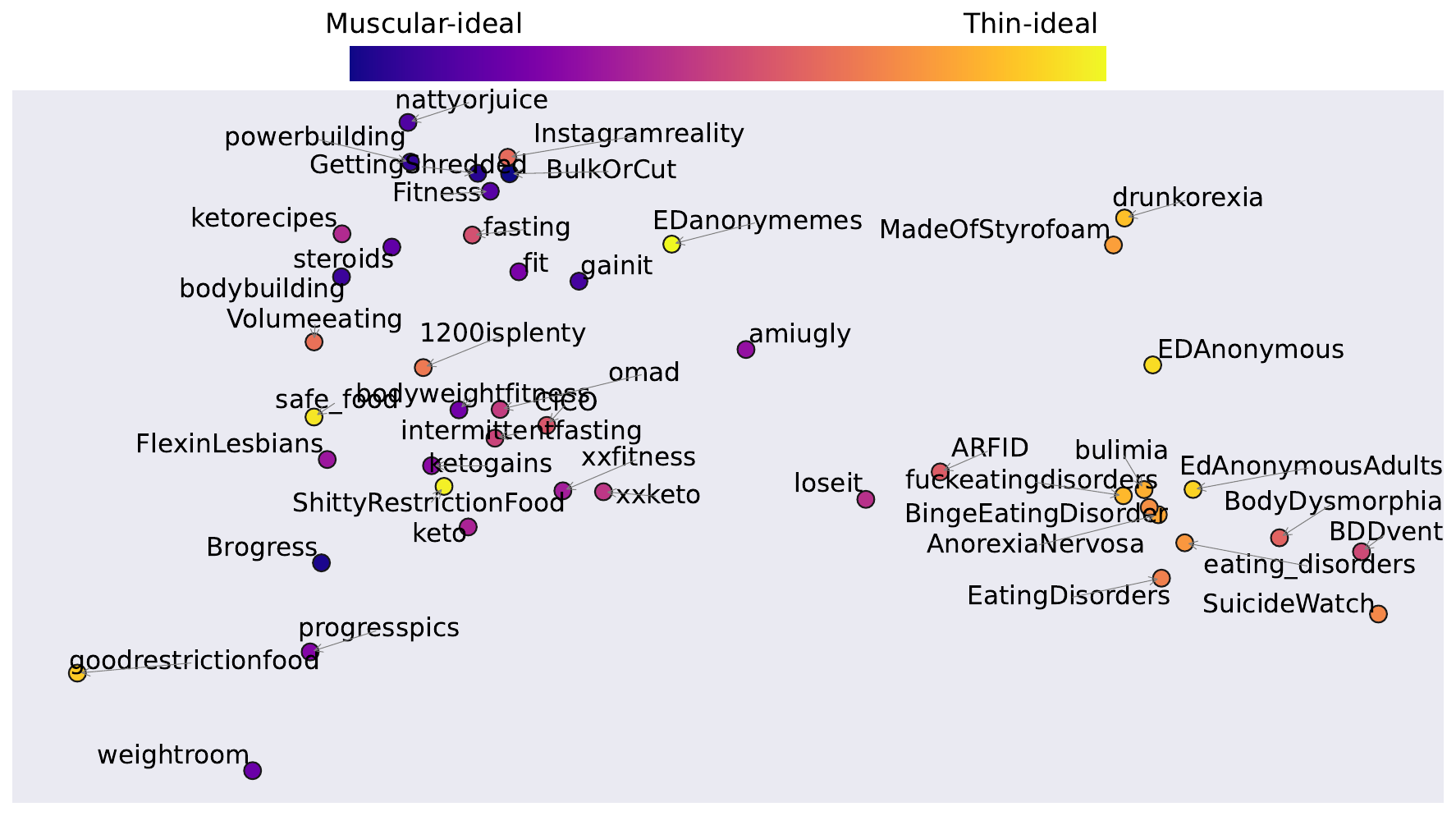}
  
  \caption{TSNE embedding of communities by their emotions and toxicity reveals the clustering of  discussions related to psychopathologies. Communities discussing mental health issues like suicidal ideation, body dysmorphia, self-harm, and eating disorders show similar emotional patterns that are distinct from other diet and fitness communities. Each community is represented by a 9d vector consisting of its 75th percentile emotion (approval, optimism, admiration, caring, disapproval, disappointment, annoyance, sadness) and toxicity score of the submissions. Communities are colored by their positions along the thin ideal-muscular ideal dimension. Communities cluster into two groups based on the affect of the submissions.
  } 
  \label{fig:emo_embeddings}
\end{figure}

\begin{figure}[ht] 
  \centering
  \includegraphics[width=\columnwidth]{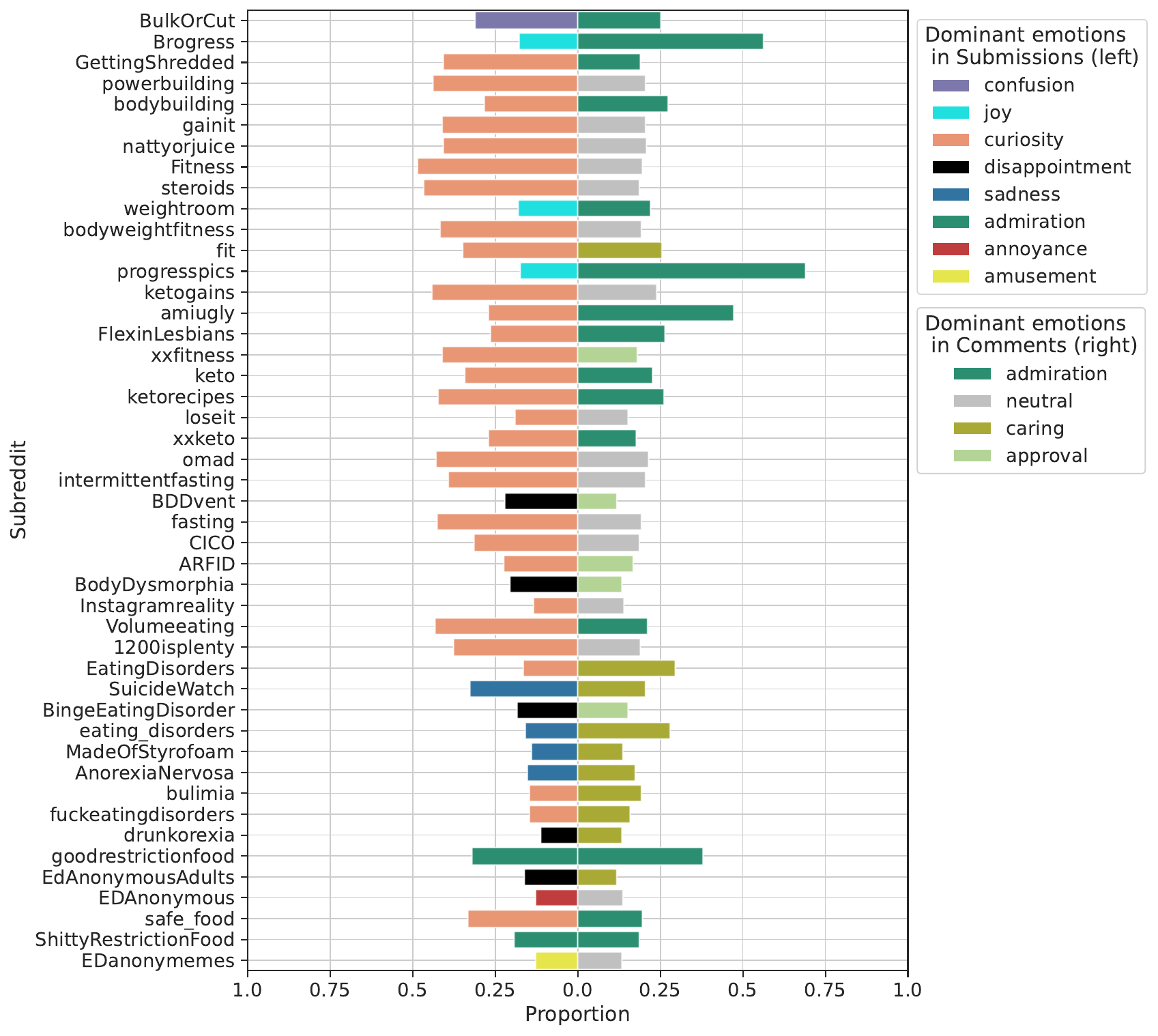}
  
  \caption{The most dominant emotion in submissions (left) and comments (right) by subreddit, ordered according to the muscular-thin ideal dimension. The analysis focuses on the top 75\% of data with the highest emotional content, excluding submissions and comments with a neutral score above the 75th percentile separately.}
  \label{fig:dom_emo_thin-ideal}
\end{figure}

\begin{figure}[ht] 
  \centering
  \includegraphics[width=\columnwidth]{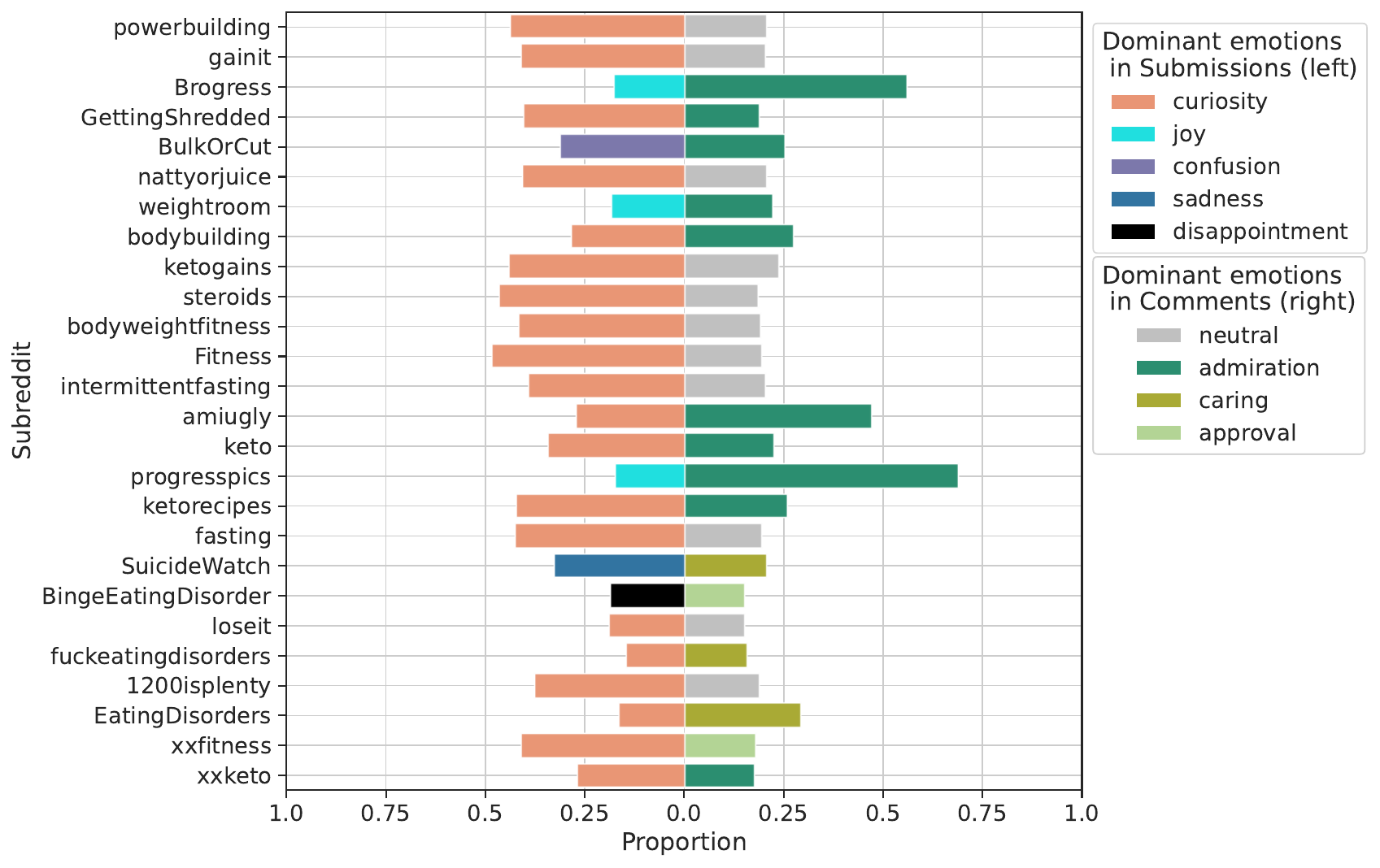}
  
  \caption{The most dominant emotion in submissions (left) and comments (right) by subreddit, ordered according to the masculine-feminine dimension. The analysis focuses on the top 75\% of data with the highest emotional content, excluding submissions and comments with a neutral score above the 75th percentile separately.}
  \label{fig:dom_emo_gender_order}
\end{figure}

Figs. \ref{fig:dom_emo_thin-ideal} and \ref{fig:dom_emo_gender_order} show the dominant emotions for different subreddits in submissions and comments according to the muscular-thin ideal dimension and the masculine-feminine dimension, respectively.

\subsection{Emotion and Toxicity Analysis of Gendered Subreddits}
\label{app:emo_tox_gendered}

\begin{figure}[ht]
  \centering
  \includegraphics[width=\columnwidth]{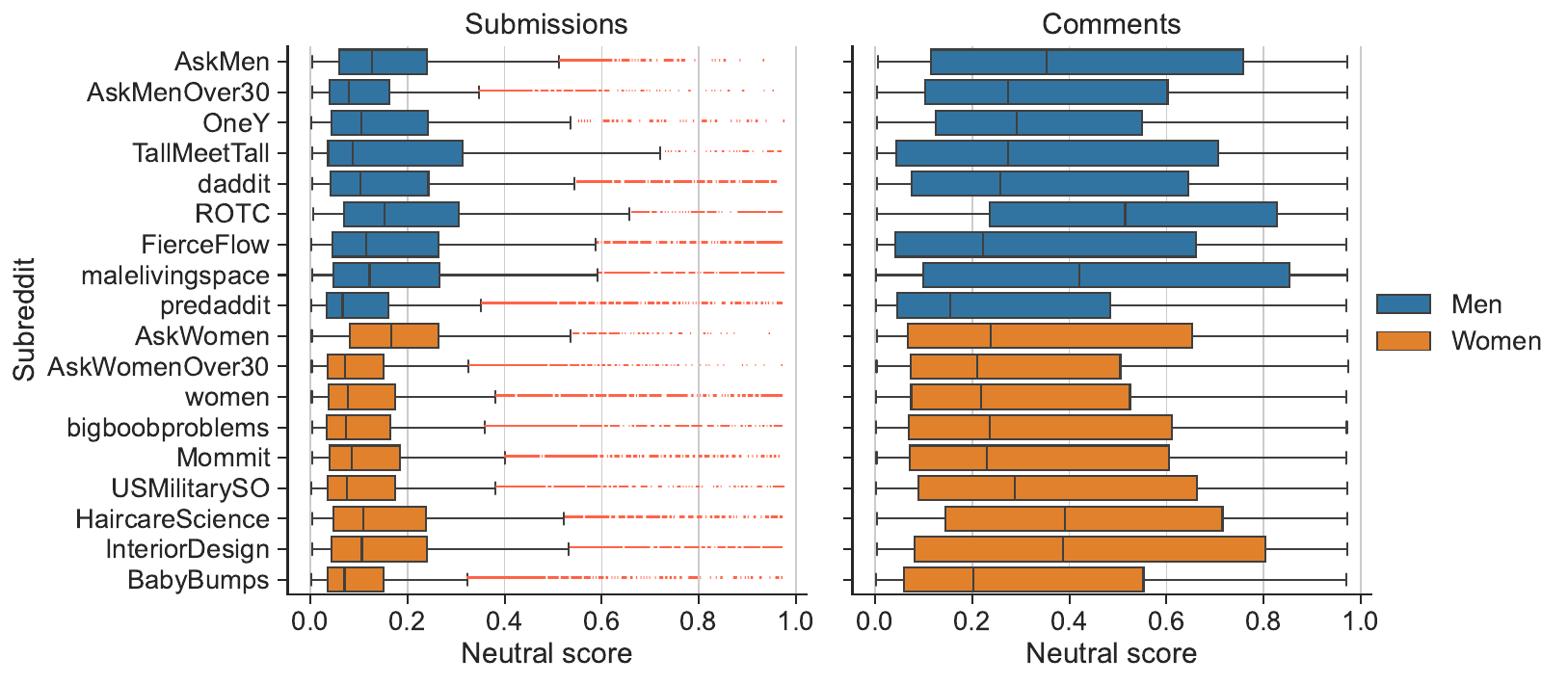}
  
  \caption{Distribution of neutral emotion scores in gendered seed subreddits.} 
  \label{fig:neutral-seed-gender}
\end{figure}

\begin{figure}[ht]
  \centering
  \includegraphics[width=\columnwidth]{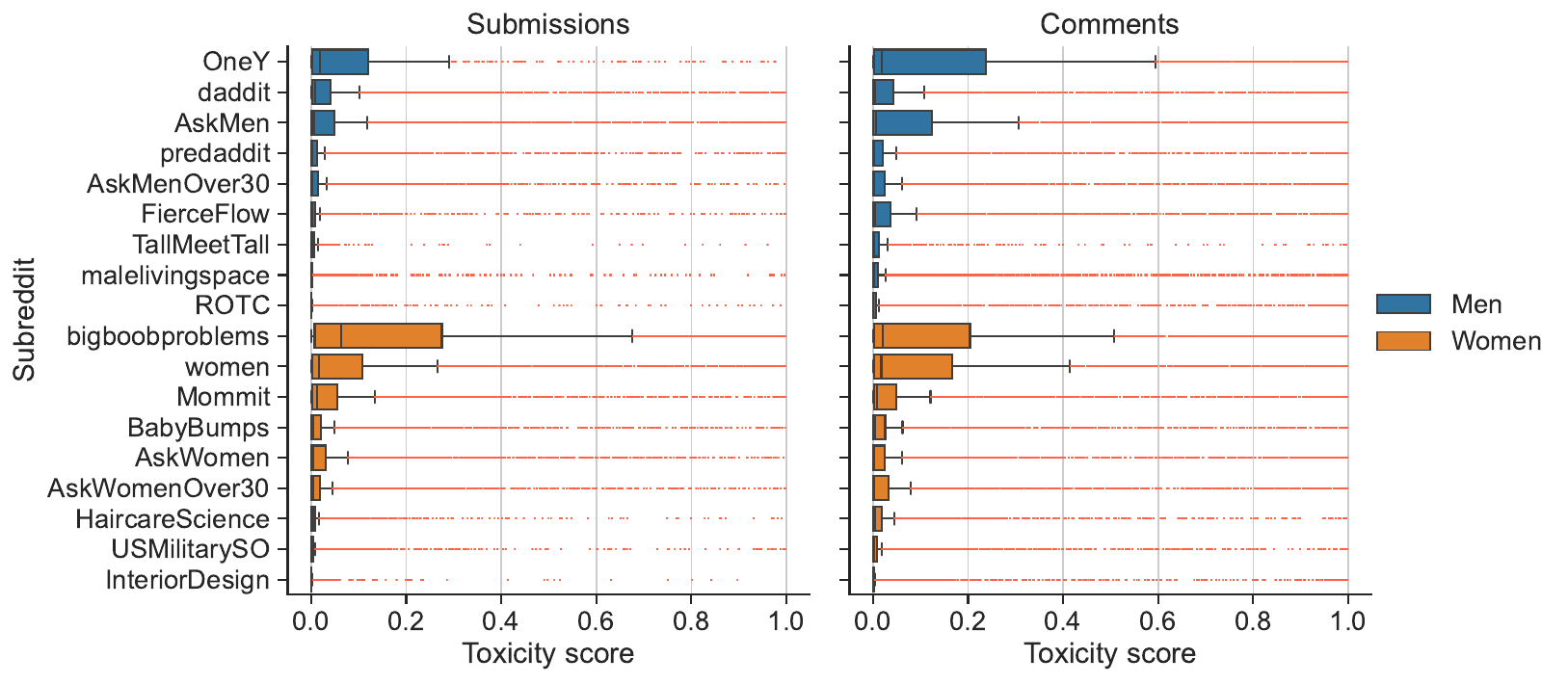}
  
  \caption{Distribution of toxicity scores in gendered seed subreddits.} 
  \label{fig:toxic-seed-gender}
\end{figure}

Figures \ref{fig:neutral-seed-gender} and  \ref{fig:toxic-seed-gender} show the distributions of neutral emotion scores and toxicity scores in seed pair communities identified by \citet{waller2021quantifying} for the gender axis.

\end{document}